\documentclass[12pt,a4paper]{article}

\begin{document}
\title{TWO MODELS OF QUANTUM BRIDGES CONNECTED WITH SEMICONDUCTORS OR METALS }
\author{V.N.Evteev, M.V.Moiseenko, E.V.Zhuravel and E.Ya.Glushko\\
Pedagogical University, Krivoy Rog 50086, Ukraine.\\ EYAGL@kpi.dp.ua}
\maketitle
\begin{abstract}
It is proposed two models describing transport and absorbtion processes that 
occur in nanoscale fragments of electrical circuits, pulled adsorbed molecules, 
atomic or molecular chains connecting electrodes. Discrete chain model of a 
molecular bridge between metallic electrodes considers quantum jumps between 
atoms containing the chain. A potential approach is represented by three-
dimensional Kronig-Penney model. The conductivity theory is developed in a 
supposition that the main contribution into the electron transfer belongs to 
non-equilibrium affinity populated states of the bridge. Current-voltage and 
thermodynamical characteristics are calculated for different cases.. 
Explanations for experimentally obtained step-like dependencies in I-V 
characteristics and its asymmetry are made. The charging effect and Coulomb 
blockade effect are discussed. It is shown the essential role of transitions 
between different bridge's  charge states. 
\end{abstract}
\newpage
\section{I. Introduction}
Semiconductor wires, pulled linear molecules, point contact atomic chains 
connected with prepared metal or semiconductor surface are examples of quantum 
fragments embedded into electric circuits \cite{Rui,Wil,Dat,Ker,Glu,Evt,Evt1,Evt2}. These elements may play a role 
of main functional unites of the circuit sufficiently determining its 
properties. The most important property of such "bottle neck"-like fragments is 
that electronic structure depends strongly on small number of captured electrons 
is the subject of our investigation. Statistically quantum bridges are systems 
with alternating particles number determined by interaction with contacting 
reservoir. In recent paper \cite{Dat} the conductivity of adsorbed bistertiothene 
molecule junction (fig.\ref{P1}) fabricated by suspended micro-bridges technique 
combined with mechanically controlled break method  was studied experimentally. 
The results obtained have shown several interesting features including stepwise 
dependence of current-voltage characteristics and sometimes its asymmetry 
{\sf }relatively the sign of applied voltage.
An example of pure quantum curcuit is represented in fig.\ref{P2}a. The role of short 
pulse voltage source can play a polar molecule embedded into the closed chain. 
Quantum voltage source polarization may be controlled by resonant external 
electromagnetic field switching the molecule into long lived triplet state. The 
current pulse arising is caused by the difference of dipole moments in ground 
and excited states. The stabilization of excited state may be implemented by 
means of voltage bias between ends of the molecule (
fig.\ref{P1}). Transition process 
in excited state modifies affinity spectrum of the molecule that in turn 
influences on the electron transport through the molecular bridge.
Our results show that the voltage applied modifies significantly the electronic 
structure of all kinds of bridges and leads to nonequilibrium redistribution of 
states population. The theory developed here for contact and transport phenomena 
in quantum bridges is based on the conception of mesoscopic system's charge 
states. Two approaches are discussed in the work for pulled quantum bridges: 
potential model and discrete chain model (
DCM). Potential model proposed 
describes semiconductor quantum wire connecting semiconductor or metal 
electrodes is shown in figure \ref{P3}.  The model is based on exactly solved problem  
for terminated Kronig-Penney crystal. 
The DCM approach describing molecular or atomic bridges starts from initial 
(atomic) generalized affinity energy $E_i $ for external reservoir electron, 
chain geometry and intersite transfer matrix elements $V_i $ (fig.\ref{P4}). Index i 
represents the number of electrons left the chain, $E_i $  is the exited 
states band center for one-fold ionized molecule. Quantum bridge capturing or 
losing electrons in absorbing process or due to applied voltage changes weakly 
its charge states $M^{(i)} $ \cite{Glu,Evt2}. General picture of charge state 
transitions for ionized quantum bridge contacting with metal surface is 
represented in the figure \ref{P3}, where $\chi $ is electrode chemical potential, 
$I_1 $ is the first ionization potential, $I_0 $ is affinity energy.
Pulled linear molecular chains adsorbed on prepared metal or semiconductor 
surface may be very interesting due to its physical properties and possible 
applications. Extremely large effective surface of these objects (Fig. \ref{P5}) in 
case its density is sufficiently high makes them attractive as adsorbing 
elements for gaseous sensors. We have proposed in \cite{Evt,Evt1} to use the forest of 
molecular chains directed by external electric field for adsorbing, storage and 
testing small concentration of gas molecules in surrounding medium. Adsorbed 
long molecular constructions are most compact and small contenders for 
conducting elements of quantum circuits, STM tips and neural networks. Meanwhile 
many properties of 1D systems absorbed in electric field on a semiconductor or 
dielectric surface demand further theoretical investigations: the conductivity 
problem at low frequencies, chain-surface local states and states arising due to 
the capture of air molecules by the chain, electron energy spectrum in external 
field, mobility of electrons, affinity spectrum, chain topology, interaction 
with vibrational modes  etc. 
In the paper we propose a theoretical consideration for some of these problems. 
Our DCM analysis is based on the electron Hamiltonian for a chain with N-sites. 
The conductivity of quantum bridges is investigated theoretically using obtained 
exact solutions for electron states of a linear chain in steady state electric 
field. It is taken into account quantum interaction between electronic subsystem 
and either metallic or semiconductor electrodes playing the role of electrons 
reservoirs. In the ground of theory proposed lies the supposition about main 
contribution of affinity states into the conductivity of point contacts both 
semiconductor wires and molecular or atomic junctions. Transfer matrix method 
was used out the framework of approximation of translation invariance and 
periodic boundary conditions. In nearest-neighbor approximation for chain atoms 
interaction and hopping probability the current-voltage characteristics are 
calculated at different temperatures and electronic structures of the chain. The 
Coulomb blockade effect is taken into account phenomenologically in a self-
consisted procedure. It is represented the explanation of current-voltage 
asymmetry observed experimentally in \cite{Dat}. It is shown the leading role of field 
modification of affinity spectrum for conductivity of molecular bridges. 
Irreversible states population and current Coulomb charging is calculated.
\section{II. An adsorbed linear molecule in electric field.}
We will describe  the problem of field influence using secondary quantization 
Hamiltonian for a trapped electron in N-periodic linear chain (Fig.\ref{P2}b) 
\begin{equation} \label{1}
\hat H_0  = \sum\limits_{k,j}^N {\varepsilon _{kj} a_{kj}^ +  a_{kj}^{}  + 
\sum\limits_{k,j,m,i}^{} {V_{km}^{ij} a_{kj}^ +  a_{mi}^{} } } 
\end {equation}
\begin {equation}
\label{2}
\varepsilon _{kj}  = \varepsilon _{0j}  + \varepsilon ekd + \varepsilon ed_l 
\end{equation}
 
where k numbers elementary cells, d marks distance between atoms along the 
chain, j numbers atoms in the elementary cell, $\varepsilon _{\rm j} $ 
determine initial atomic affinity levels, the hermitian matrix   $V_{nm}^{ij} 
$ describes interatomic electron transfer, $d_l $  is the length of 
absorbtion bonds on the left-hand side of linear molecule, $\varepsilon $ 
represents applied electric field. The matrix of Hamiltonian (\ref{1}) will be wrote 
in nearest neighbour approximation for an adsorbed molecular chain like 
polyacetylene $R - (CH)_n  - R$. In the model under consideration the 
transfer between elementary cells is possible through carbon atoms only; the 
amplitude of the process equals W, hopping amplitudes inside elementary cells 
equals V ( Q is its value for end atoms). Considering $U = \varepsilon (Nd - d 
+ d_l )$ as the whole voltage between chain ends we can obtain the energy 
dipersion equation 
\begin{equation}
\label{3}
(\lambda _l , - \nu _l )\prod\limits_{k = 2}^{N - 1} {\hat \Lambda _k } \left( 
{\begin{array}{*{20}c}
   {\nu '_r }  \\
   {\lambda '_r }  \\
\end{array}} \right) = 0
\end{equation}
\[
\hat \Lambda _k  = \left( {\begin{array}{*{20}c}
   {\mu _k } & {\nu _k }  \\
   {\lambda _k } & 0  \\
\end{array}} \right)
\]
containing transfer matrix product instead of transfer matrix power. Here
\[
\mu _k  = (\varepsilon _0  - E + \varepsilon ekd + \varepsilon ed_l 
)(\varepsilon _1  - E + \varepsilon ekd + \varepsilon ed_l ) - V^2 
\]
\[
\nu _k  =  - \lambda _k  = W \cdot (\varepsilon _0  - E + \varepsilon ekd + 
\varepsilon ed_l )
\]
\[
\nu '_r  = (\varepsilon _0  - E + U - \varepsilon ed_r )((\varepsilon _0  - E + 
U - \varepsilon ed_r )(\varepsilon _1  - E + U - \varepsilon ed_r ) - 2Q^2 
\]
\begin{equation}
\label{4}
\lambda _r^{'}  =  - W \cdot (\varepsilon _0  - E + U - \varepsilon ed_r )^2 
\end{equation}
Using canonical transformation and the procedure described in \cite{Glu} the equation 
(\ref{3}) is transformed to convenient view obtained in absence of external field.
\begin{equation}
\label{5}
e_{} ^{f_1 } (\nu _r y_{22}  - \lambda _r y_{21} )(\lambda _0 y_{11}  - \nu _0 
y_{12} ) - e_{} ^{f_2 } (\nu _r y_{12}  - \lambda _r y_{11} )(\lambda _0 y_{21}  
- \nu _0 y_{22} ) = 0
\end{equation}
where $f_1$ and $f_2 = - f_1$   are eigenvalues of matrix $hat F$ in 
the power of exponent 
\[
\hat F = Ln(\prod\limits_k {\hat \Lambda _k )} 
\]
\begin{equation}
\label{6}
\prod\limits_{k = 2}^{N - 1} {\hat \Lambda _k }  = \exp \left( {} \right.\sum\limits_k {\left( {\begin{array}{*{20}c}   {\frac{{\lambda _1^k \ln \lambda _1^k  - \lambda _2^k \ln \lambda _2^k }}{{\lambda _1^k  - \lambda _2^k }}} & {\frac{1}{{\lambda _2^k  - \lambda _1^k }}\ln \frac{{\lambda _2^k }}{{\lambda _1^k }}}  \\
   {\frac{1}{{\lambda _2^k  - \lambda _1^k }}\ln \frac{{\lambda _2^k }}{{\lambda _1^k }}} & {\frac{{\lambda _1^k \ln \lambda _1^k  - \lambda _2^k \ln \lambda _2^k }}{{\lambda _2^k  - \lambda _1^k }}}  \\ \end{array}} \right)} \left. {} \right)
\end{equation}
Here $ \lambda _i^k  = (\mu _k^{}  + (\mu _k^2 + 4\lambda _k \nu _k )^{1/2})/2 $ are eigenvalues of transfer matrix taking part in the product. The 
condition $ \mu _k^2  + 4\lambda _k \nu _k  = 0$ determines imagine boundary 
between extended band states and that localised due to the influence of extended 
field. As an another matter for the states localization may serve structure 
defects including both ends of the chain. The calculations show that with the 
growth of electric field $ \varepsilon $ the extended states energy range is 
narrowed in directions from band edges to the middle. At the same time the band 
width increases to the value of applied voltage U. The last extended states 
transforms to localized ones when U exceeds starting band width. The 
distribution of the electron density amplitudes along the chain is given by 
eigenvectors of the problem solved above. The model under consideration (DCM) 
allows to obtain analytical expressions for eigenvectors determined by 
coefficients $ C_{kj}$ of the canonical transformation 
\[
a_s^ +   = \sum\limits_{k,j} {C_{kj} (s)} a_{kj}^ +  
\]
\[
a_s^{}  = \sum\limits_{k,j} {C_{ij}^* (s)} a_{kj}^{} 
\]
The summation is performed on chain cells k and on atoms j inside the elementary 
cell. Index s=1,2,...N numbers in the work the electron affinity states. One can 
obtain the coefficients $C_{ij}$ analytically using well known Kramer's rule and 
revealing determinants corresponded to each variable.
\begin{equation}
\label{7}
C_{k1}^{}  = d_{k - 1}  \cdot QV\prod\limits_{i = k + 1}^N {( - \varepsilon _{0i} W)} 
\end{equation}
\[
C_{k2}^{}  = C_{k1}^{} \frac{V}{{E_s  - \varepsilon _0 }}
\]
\begin{equation}
\label{8}
d_{k - 1}  = \frac{1}{{\Theta _{k - 3} }}(e_{} ^{f_{k - 3} } (\nu _r y_{22}  - \lambda _r y_{21} )(\lambda _0 y_{11}  - \nu _0 y_{12} ) - e_{} ^{ - f_{k - 3} } (\nu _r y_{12}  - \lambda _r y_{11} )(\lambda _0 y_{21}  - \nu _0 y_{22} ))
\end{equation}

  The determinants of k-order $\Theta _k $ are determined by the left upper edge 
of the Hamiltonian dynamic matrix described in (\ref{1}). Numerical calculations for 
electron density $|C_{kj} (s,i)|^2 $ both with the account of field dependence 
in transfer matrix elements V and without that influence was performed in \cite{Glu}. 
With the growth of applied voltage the standing probability waves $|C_{kj} 
(s,i)|^2 $ become less symmetric (or anti-symmetric) respectively the center of 
the linear molecule. At the same time, the interatomic barriers begin to 
decrease at sufficiently great fields ($~10^9$ V/m ) that leads to repairing of 
the symmetry in electron density distribution.
\section{III. Electronic specific heat of adsorbed molecules.}
The contribution of affinity electrons captured by a free molecule into the 
specific heat $C_{aff}$ is determined by expression
\begin{equation}
\label{9}
C_{aff}  = T{}^{ - 2} \cdot \sum\limits_s {[(E{}_s - E{}_1){}^2n{}_s(1 - 
n{}_s)]} 
\end{equation}
where $n{}_s = (\exp (\frac{{E_s  - E_1 }}{T}) + 1){}^{ - 1}$, $E_1$ marks the lowest level that plays the role of system chemical 
potential. The temperature dependence of $C_{aff}$ calculated by (\ref{9}) in absence of 
electric field for a long 20-atomic chain is represented in fig.\ref{P6}, curve 1. 
Energy parameters of the chain were taken  $\varepsilon _0 $=-4,3 eV, V=0,02 eV 
. In case of an adsorbed molecule its specific heat $C_{aff}$  depends on affinity 
band position relatively the surface chemical potential. We suppose here that 
the latter coincides with band center $\chi _1  = \varepsilon _0 $. In case all 
levels lying below $\chi_1$ the states are filling completely by left reservoir 
electrons. The curve 2 in fig.\ref{P6} shows $C_{aff}$ temperature dependence for 
molecules adsorbed in external electric field U=0.5 V/molecule on a gold surface 
$\chi_1 $.=-4,3 eV calculated by (\ref{6}) at $E_1=\chi_1$ . We do not take into 
account the Coulomb blockade effect in specific heat calculations.
        It should be marked that external (negative) electric field is 
significant stabilizing factor, which allow linear molecule to contact with 
adsorbing surface by left or right end only. Field absence may cause the capture 
of a molecule by the surface due to attracting image forces with molecule 
following reconstruction. Negative electric field interacting with dipole 
momentum arising in chain due to its charging pulls molecule normally to the 
surface (fig.\ref{P5}). Opposite field direction vice versa overturn the molecule on 
adsorbing surface. Changing electric field may influence strongly on molecular 
strain and orientation that is on the effective volume of molecular forest. 
To compare we consider the same chain connecting two gold electrodes having 
different potentials. At room temperature T=0.025 eV calculations give falling 
$C_{aff}$dependence on applied voltage U (fig.\ref{P6}, curve 3). The maxims observed 
are well known thermodynamic Shottky anomalies. 
\section{IV. Semiconductor quantum wire in electric field.}
To describe contact and transport phenomena in semiconductor quantum wires we 
used exactly solvable 3D Kronig-Penney model with open boundary conditions. The 
proposed potential model describing quantum bridge with sizes 20x20x100 
elementary cells is based on exact solution for terminated 1D Kronig-Penney 
crystal with $\delta$-functional barriers (Dirac'comb potential) obtained in 
\cite{Glu1} without using of translation invariant approximation. One-dimensional model 
crystal contains N wells of width a, deep $U_0$ and opaque coefficient  $\Omega 
$. As a result of superposition of the same potential in three dimensions it is 
obtained a separable 3D potential shown in fig.\ref{P7}. The summation of superposing 
combs creates suitable system of 3D wells and barriers inside the crystal volume 
but there arises simultaneously a potential   $\Delta U(x,y,z)$  additional to 
crystal one. In different ranges $\Delta U(x,y,z)$ takes values $U_0$, $2U_0$  
and combinations of $U_0$ and alternating barriers $\Omega$ (fig.\ref{P7}). The 
Hamiltonian of the problem may be represented as
\begin{equation}
\label{10}
\hat H = \hat H_0 (x,y,z) - \Delta U(x,y,z),
\end{equation}
\[
\hat H_0 (x,y,z) = \hat H_{0x}  + \hat H_{0y}  + \hat H_{0z} 
\]
where $H_0$ is separable part of entire Hamiltonian. The main idea of the method 
proposed is based on the fact that corrections to zero-order results are small 
because of wave function's tails in additional ranges of   $\Delta U(x,y,z)$ are 
asymptotically small for band states and to some extent for deep local states. 
Zero wave function is represented in a view of product 
\[
\tilde \Psi _0 (x,y,z) = \tilde \Psi _{0x} (x)\tilde \Psi _{0y} (y)\tilde \Psi 
_{0z} (z)
\]
where  $\tilde \Psi _{0x} (x),\;\tilde \Psi _{0y} (y)\;,\tilde \Psi _{0z} (z)$ 
are 1D solutions obtained exactly in \cite{Glu1}. Non-additive addition leads to 
asymptotically small corrections for band states. We have studied GaAs and 
AlGaAs quantum bridges between metallic electrodes using active computer 
designer of hierarchical structures (ACDHS) which allows to built and to 
calculate different potential well systems both periodical and hierarchical.
Electric field modifies states of semiconductor quantum bridge similar to 
considered above molecular chains. Let we consider GaAs quantum bridge of length 
199,15 $\mathop A\limits^0 $ connecting metallic electrodes in steady state 
electric field with intensity $\varepsilon$. The Kronig-Penney crystal 
parameters in case of GaAs satisfying to its affinity energy are $U_0$=1.2278 
eV, $\Omega a$ = - 0.215 , a=5.69 $\mathop A\limits^0 $. Taking into account the 
GaAs band gap value $E_g$=1.52 eV we will consider voltage bias small enough to 
avoid interband transfers$e\varepsilon L \ll E_g $ , where L is the crystal 
length. Calculations by means of ACDHC give band structure, band position and 
distribution of electron density in different exited states. Figure \ref{P8}a shows the 
system geometry and relative positions of energy bands in semiconductor wire and 
in metal. Figures\ref{P8} b and \ref{P8}c represent electron density distribution near the 
band bottom and top, respectively. The field applied manifests itself in states 
modification leading to electron density drive away from left-hand or right-hand 
shores.
\section{V. Conductivity of quantum bridges.}
Theoretical analyze of bridge conductivity is based mainly on two approaches. 
Landauer wave model \cite{Ree,Avi} interprets the molecular junction as an scattering 
center reflecting electronic waves moving from cathode. Electrical current is 
proportional either to transmission coefficient of the electronic wave near the 
Fermi energy of the electrode or to some integral of the transmission in a 
proper energy range \cite{Dat,But}. Calculations of linear conductance in wave model 
give a more intensive current in comparison with experimental data \cite{Dat,Ree,Avi}. 
As it was shown in \cite{Glu} this effect is the consequence of both neglecting by 
spectrum field modification and by important role of molecular affinity charge 
states. The kinetic (sequentional) conductivity model \cite{Dat,Emb,Por} operates by a 
phenomenological transfer rate describing threefold tunneling processes of 
electron's jumps between a cathode and a couple of molecular states. Though 
calculations of conductance in kinetic model for gold-bistertiothene-gold 
contact \cite{Dat} gave a good agreement with the experiment as to the values order and 
reflects main features of current voltage dependencies, we suppose the 
theoretical grounds of the model leaves much to be desired due to the 
independent physical meaning of threefold processes. We will consider below the 
conductivity of molecular bridge  in DCM. The bridge electron subsystem relation 
with the electrodes manifests itself in nonzero possibility amplitude of 
electron transfer onto the molecule. Respective addition to the Hamiltonian 
\begin{equation}
\label{11}
\hat H_{ad}^{}  = G_l^{}  \cdot (a_l^ +  a_{10}^{}  + a_{10}^ +  a_l^{} ) + 
G_r^{}  \cdot (a_r^ +  a_{N,0}^{}  + a_{N,0}^ +  a_r^{} )
\end{equation}
contains jump amplitudes $G_l$,r between metallic shores and end atoms of the 
chain. Indexes l,r correspond to left and right shores, respectively. In case of 
an adsorbed molecule (\ref{11}) contains only one term. The model presented determines 
electronic structure of metallic electrodes by density of states
\begin{equation}
\label{12}
g(\varepsilon _{r,l} ) = \frac{{4\pi {\rm H}}}{{h^3 }}(2m)^{3/2} \varepsilon 
_{r,l}^{1/2} 
\end{equation}
where H is effective volume of the electrode contacting the molecule, m is 
electron effective mass, $\varepsilon_{r,l}= E_{Frl} - 
\chi_(r,l)+E_{rj}+U_{l,r}$ expresses the state energy $E_{sj}$, measured from 
the beginning of electrode's Fermi-stage with account of the bias value 
$U_{l,r}$, $E_{Frl}$ marks right or left Fermi-energy. Both chemical potential 
$\chi_x,l$ and energy $E_{sj}$ are measured respective to vacuum level accepted 
to be zero (Fig.\ref{P4}). Due to the field widening of the discrete energy band the 
affinity levels pass sequentially the active ranges near Fermi surfaces of both 
shores. Each pass corresponds to the current sharp increasing in I-V dependence. 
There is a current dependence on the density of states both left-hand and right-
hand electrodes. This effect is significant one in case of semiconductor shores. 
Intrinsic phonons (or vibrons) influence weakly on the electron transfer and 
non-equilibrium states population putting in a molecule with current \cite{Glu}. The 
shores play the role of infinite thermodynamically balanced reservoirs of 
electrons. Equilibrium populations of baths'electronic states are described by 
the Fermi distribution. In model under consideration the injected electron 
occupies either one of the affinity states $E_{sj}$ or one of vibrational 
sublevels. The possibility rate of the process can be expressed in second order 
of perturbation theory using well known Fermi golden rule. We suppose the weak 
absorbtion contact plays the role of a small parameter. We will consider the 
phononless contribution into the conductivity. Then one can obtain from the 
coinciding for left end and right end electronic currents through bridge's sj-
state the entire current I through the junction
\begin{equation}
\label{13}
\begin{array}{l}
 I = \frac{{2\pi e}}{\hbar }\sum\limits_{s,j} {} |G_{l,sj} |^2 |G_{r,sj} |^2 g_l 
(E_{sj}  - \chi _l  + E_{Fl}  + U_l ) \cdot g_r (E_{sj}  - \chi _r  + E_{Fr}  + 
U_r ) \cdot  \\ 
 {{(N_l  - N_r) }} / {{(|G_{l,sj} |^2 g_l (E_{sj}  - \chi _l  + E_{Fl}  + U_l ) + 
|G_{r,sj} |^2 g_r (E_{sj}  - \chi _r  + E_{Fr}  + U_r ))}} \\ 
 \end{array}
\end{equation}
where the summation is performed over all affinity states. The dimensional part 
of entire current 
\begin{equation}
\label{14}
I_0  = \frac{{e \cdot (2m)^{3/2} {\rm H}G^2 }}{{\pi \hbar ^4 }} \cdot 
(1eV)^{5/2} 
\end{equation}
where G is the amplitude of the shore-bridge transfer, plays the role of a 
current unity. Numerical value $I_0$, in case all energies in (\ref{13}) are measured 
in eV and the effective volume H is taken in $\mathop {A^3 }\limits^0 $, is 
equal to 10.41x$HG^2$ mkA. The expression obtained takes into account both the 
difference between materials of left and right electrodes and possible asymmetry 
of voltage connection in the circuit relatively the vacuum level. In the 
symmetric case 
\[
U_r=-U_l=U/2 ,
\]
then in case of similar metallic electrodes the current-voltage characteristics 
is symmetric one relatively the applied voltage. In case the left electrode is 
earth then one should take in (\ref{13}) $U_l =0$ and $U_r=U$ . Therefore, the I-V 
dependence has lost its symmetry relatively the sign of applied voltage. In 
experimental work \cite{Dat} it was presented the examples both symmetric and 
asymmetric I-V characteristics of quantum junctions. Our calculations show that 
the current value vs voltage depends sufficiently on the distribution of 
electron densities near ends of the chain$|C_{sj} (l,0)|^2 $ and $|C_{sj} 
(N,0)|^2 $ in the standing wave as well as from the difference in metallic 
shores populations on the absolute energy level $E_{sj}$. The influence of 
square root energy dependence of state densities $g_l$ and $g_r$ manifests 
itself not so strongly. An essential fact for understanding the phenomenon of 
current going through a mesoscopic linear system is that that characteristic 
temperatures being of order $T\sim 0.02-0.03 eV$ are as a rule much less than the 
distances between levels of bridge  affinity band. Therefore with growth of 
applied voltage U the affinity levels one by one pass through the range near the 
Fermi surfaces EFl and $E_{Fr}$ where the transfer become very effective. The 
entry of each new level into the active range accompanies by sharply increasing 
of current. The energy pauses between levels cause a plateau I-V dependence. The 
charging - entire charge trapped by chain has similar behaviour. Below we will 
analyze taking, for example, a simple chain, the nature of electron transfer 
through the molecular bridge connecting metal electrodes without taking into 
account the Coulomb blockade effect.  The contact conditions may be so that the 
external electric field is negligible small at comparatively big voltages. It is 
the case when a pulled molecule connects the tips of fine electrodes and the 
drop of voltage occurs in vicinity of molecular ends. External electrical field 
doesn't influences the molecular affinity states. The solid line in Fig. \ref{P9} shows 
calculated I-V characteristics for a chain with four starting affinity states at 
$E_0=$-4.0 eV,  V=0.2 eV, $U_0=$ 0.5 eV, E=0.01eV in case of gold electrodes. 
Dotted line  shows  the  charging effect.  One can see  the characteristic  
tendency of  fraction and semi-nteger charging the molecule under voltage as the 
matter of thermodynamically non-equilibrium states population. The almost 
exactly semi-integer chain's charge arises at sufficiently low temperatures 
T<<V, if the energy range $(\chi,\chi+U)$ contains an odd number of affinity 
states \cite{Glu}. States laying above the range are unoccupied ones $n_s=0$ , down 
states are filled completely $n_s=1$  . Immediate states marked in case of 
symmetric contacts molecule-electrodes have populations $n_s"1/2$. An additional 
circumstance important for semi-integer charging effect is approximate equality 
for electrodes' density of states $g_r"g_l$ (\ref{12}) small bias U under 
consideration.
Many body character of the problem may be taken into account phenomenologically 
in a self-consisting procedure including general spectrum shift due to the 
charging and band widening. We have used for ground state energy $\varepsilon_i$ 
in $M^{(i)}$-state.
\begin{equation}
\label{15}
\varepsilon_i=\varepsilon_{i+1}+\varphi _i  - 2V_{i + 1} \exp (\varphi _i 
/\varphi _{i + 1} )
\end{equation}
Coulomb barrier potentials $\varphi_i$ are determined in a minimization 
procedure for many electron system of the proper state $M^{(i)}$ Starting from 
intial charge state $M^{(i)}$ we have for bridge's many-body energy terms 
\begin{equation}
\label{16}
\begin{array}{l}
 X_1  = \varepsilon _1  - 2V_1  = E_v  \\ 
 X_0  = 2(\varepsilon _0  - 2V_0 ) \\ 
 X_1  = 3(\varepsilon _{ - 1}  - 2V_{ - 1} ) \\ 
 X_2  = 4(\varepsilon _{ - 2}  - 2V_{ - 2} ) \\ 
 \end{array}
\end{equation}
where $E_v$ is absolute valence band top position of neutral molecole. Using 
(\ref{15}) we obtain from (\ref{16}) the ionization potentials of the chain  under 
consideration 
\[
I_i=X_i-X_{i-1}
\]
Calculations performed in DCM framework show a good agreement with experimental 
data for ionization potentials of carbon nanosystems and molecular wires I-V 
characteristics. 
\section{VI. Conclusion.}
        Quantum bridges under the current are parametric many-body systems 
having steady state non-equilibrium distribution. The number of electrons 
captured by the molecule's affinity states from outside is controlled by 
external conditions that determine the molecule charge's state described by own 
electronic structure. The discrete chain model considered as well as Kronig-
Penney model belong to the number of simplest ones. Nevertheless it allow to 
take into account in an united formalism the main features of current carrying 
through the mesoscopic junctions. The cathode electrons at first occupy the 
system eigenstates and then they transit to anode. The model supposes existing 
an initial affinity level in each potential well. Due to the finite barriers 
width there arises a band of states in the system of potential wells. The band 
structure and its situation relatively the shores chemical potentials determines 
the current character in framework of given molecular charge's state. Following 
charging creates a Coulomb barrier modifying the affinity spectrum. The electron 
spectrum of neutral molecule $M^0$ begin participate in conductivity in case of 
a single-ionized molecule. These states fills partially by an electron transited 
from the electrode. With the growth of applied voltage total population may 
exceed unity. In the case it is switch on the next affinity spectrum $M^l$ and 
so on. In present work the Coulomb charging was taken into account 
phenomenologically. A more correct approach demands the resolution of a quantum 
many-body problem for discrete chain with alternative number of particles 
depending on voltage applied. 
Both models able to describe low-temperature electron transfer in metallic point 
contacts. The structures were investigated experimentally in \cite{Rui,Yan} for 
monoatomic wires between Au, Pt, Al, Nb, Pb, K è Na electrodes. The transfer 
mechanism through affinity states leads also for such systems to stepwise 
conductance with both positive and negative step slopes. Our models give an 
other alternative to explain the vanishing of conductivity at sufficiently long 
distances between electrode's tips. The matter may be laying in electronic 
structure transformation lifting due to the Coulomb blockade effect the bottom 
of conductivity band upper than Fermi level. 
Quantum fragments of electric circuit plays important role in single electron 
transistors and ratchets \cite{Wan}. Though the linear models considered can't be 
directly applied to such systems, one should mark the general equivalence 
between  bridges and quantum dots possessing discrete spectrum and contacting 
with several electrodes.

\newpage
\mbox{}
\begin{figure}

\setlength{\unitlength}{0.1pt} 
\begin{picture} (354,149) (0,0)
\special{em:graph 1.gif}
\end{picture}
\vspace{200pt}
\caption{BISTERTIOPHENE MOLECULE AS A QUANTUM FRAGMENT 
OF ELECTRIC CIRCUIT INVESTIGATED EXPERIMENTALLY 
IN \cite{Dat}}

\label{P1}
\end{figure}

\newpage
\mbox{}
\begin{figure}  
\setlength{\unitlength}{0.1pt}
\begin{picture}(00,50)(-450,0)
\special{em:graph 2.gif}
\end{picture} 
\vspace{160pt}
\caption{A DIPOLE MOLECULE AS AN EMF SOURCE IN AN ELECTRIC CIRCUIT \newline
a) Complete electric circuit with an inverse asymmetric molecule as voltage source 
activated by light. DP is the dipole momenta difference in ground and excited states. 
Zigzag show light causing dipole changing.\newline
b) A model of hydrocarbon linear molecule. W is transfer amplitude between carbon 
atoms, V is the same inside the elementary cell; Q is the same for end cells.}
\label{P2}
\end{figure}

\newpage
\mbox{}
\begin{figure}[t]  
\setlength{\unitlength}{0.1pt}
\begin{picture}(406,289)(0,0)
\special{em:graph 3.gif}
\end{picture} 
\vspace{300pt}
\caption{SEMICONDUCTOR QUANTUM WIRE \newline
a,c mark electrodes, b represents quantum wire, elementary cells are shown by lines.}
\label{P3}
\end{figure}

\newpage
\mbox{}
\begin{figure}  
\setlength{\unitlength}{0.1pt}
\begin{picture}(100,200)(-450,0)
\special{em:graph 4.gif}
\end{picture} 
\vspace{290pt}
\caption{A SCHEME OF QUANTUM BRIDGE CHARGE STATES \newline
$M^{(i)}$ mark quantum bridge charge state with i electrons left the bridge,
$I_i$ corresponds to ionization energy of respective state.}
\label{P4}
\end{figure}

\newpage
\mbox{}
\begin{figure}  
\setlength{\unitlength}{0.1pt}
\begin{picture}(100,200)(-250,0)
\special{em:graph 5.gif}
\end{picture} 
\vspace{280pt}
\caption{THE FOREST OF LINEAR MOLECULES ADSORBED ON CONDUCTING SURFACE\newline
Molecular chain forest on the substrate. Dark colour marks an absorbed gas molecule.}
\label{P5}
\end{figure}

\newpage
\mbox{}
\begin{figure} 
\setlength{\unitlength}{0.1pt}
\begin{picture}(100,200)(-450,0)
\special{em:graph 6.gif}
\end{picture} 
\vspace{240pt}
\caption{SPECIFIC HEAT TEMPERATURE DEPENDENCE OF A LINEAR CHAIN \newline
Curves 1, 2 show specific heat for free molecule and for adsorbed molecule,
 respectively (lower axe). Curve 3 (upper axe) is calculated $C_{aff}$ for a 
 molecular bridge under current.}
\label{P6}
\end{figure}

\newpage
\mbox{}
\begin{figure}  
\setlength{\unitlength}{0.1pt}
\begin{picture}(100,200)(-450,0)
\special{em:graph 7.gif}
\end{picture} 
\vspace{200pt}
\caption{POTENTIAL MODEL OF A QUANTUM WIRE CONNECTING METALLIC CONTACTS
\newline
The range 1 is additional potential at parallelepiped tops, range 2 is additional
 potential at ribs, intrinsic range 3 is semiconductor wire and range 4 represents 
 metallic electrodes.}
\label{P7}
\end{figure}

\newpage
\mbox{}
\begin{figure}  
\setlength{\unitlength}{0.1pt}
\begin{picture}(100,200)(-450,0)
\special{em:graph 8.gif}
\end{picture} 
\vspace{250pt}
\caption{QUANTUM WIRE IN ELECTRIC FIELD \newline
(a) chain of 35 GaAs potential wells at applied voltage
 $dU=0$,152 eV, $\chi$ is chemical potential of 
 electrodes,$a=$5.69 \AA , 
$\Omega a=$-2.1337, $U_0=$1.2278 eV, 
dotted line marks the level of vacuum. (b, c) electron 
wave functions of the lower 
and upper exited states s=3, s=35 respectively.}
\label{P8}
\end{figure}

\newpage
\mbox{}
\begin{figure}  
\setlength{\unitlength}{0.1pt}
\begin{picture}(100,200)(-450,0)
\special{em:graph 9.gif}
\end{picture} 
\vspace{200pt}
\caption{. A QUANTUM BRIDGE I-V CHARACTERISTICS \newline
$E_0=$-4.0 eV, $V_0=$0.2 eV,$ U_0=$0.08 eV. Solid line represents current-voltage 
dependence without charging for intermediate electronic band at T=116 K; 
Dotted line shows trapped charge.}
\label{P9}
\end{figure}

\end{document}